\documentclass[prb,aps,twocolumn,superscriptaddress]{revtex4-1}
\usepackage{graphicx}
\usepackage{dcolumn}
\usepackage{bm}
\usepackage{bbm}
\usepackage{amsfonts}
\usepackage{amsmath}
\usepackage{natbib}
\usepackage{braket}
\usepackage{fixltx2e}
\usepackage[normalem]{ulem}
\providecommand{\keywords}[1]{\textbf{\textit{Index terms---}} #1}
\usepackage[mathlines]{lineno}
\usepackage[colorinlistoftodos]{todonotes}
\usepackage[colorlinks=true, allcolors=blue]{hyperref}

\begin{document}

\newcommand{\unamone}{Departamento de Sistemas Complejos, Instituto de Fisica,
Universidad Nacional Aut\'onoma de M\'exico, Apartado Postal 20-364,01000,
Ciudad de M\'exico, M\'exico.}
\newcommand{\unamtwo}{Instituto de F\'isica,
Universidad Nacional Aut\'onoma de M\'exico, Apartado Postal 20-364 01000,
Ciudad de  M\'{e}xico, M\'exico}
\newcommand{\uabc}{Facultad de Ciencias, Universidad Aut\'onoma de Baja California, Apartado Postal 1880, 22800 Ensenada, Baja California, M\'exico}.

\title{Electron transitions for Dirac  Hamiltonians with flat-bands  under electromagnetic radiation and its application to the $\alpha-\mathcal{T}_3$ graphene model}

\author{M. A. Mojarro}
\affiliation{\uabc}
\author{V. G. Ibarra-Sierra}
\email{vickkun@fisica.unam.mx}
\affiliation{\unamone}
\author{J. C. Sandoval-Santana}
\affiliation{\unamtwo}
\author{R. Carrillo-Bastos}
\affiliation{\uabc}
\author{Gerardo G. Naumis}
\affiliation{\unamone}

\date{\today}

\begin{abstract}
In a system with a Dirac-like linear dispersion there are always states that fulfill the resonance condition for electromagnetic radiation of arbitrary frequency $\Omega$. When a flat band is present two kinds of resonant transitions are found. Considering the $\alpha-\mathcal{T}_3$ graphene model as a minimal model with a flat band and Dirac cones, and describing the dynamics using the interaction picture, we study the band transitions induced by an external electromagnetic field. We found that transitions depend upon the relative angle between the electron momentum and the electromagnetic field wave-vector. For parallel incidence, the transitions are found using Floquet theory while for other angles  perturbation theory is used.  In all cases, the transition probabilities and the frequencies are found.  For some special values of the parameter $\alpha$ or by charge doping, the system behaves as a three-level or a two-level Rabi system. All these previous results were compared with  numerical simulations. A good agreement was found between both. The obtained results are useful to provide a quantum control of the system.

\end{abstract}
\maketitle
\section{Introduction}
Following the recent discovery of correlated insulation and avowedly unconventional superconductivity in twisted bilayer graphene \cite{Cao2018}, the interest on the physics related to this system, and in particular to the flat bands that appears at the so-called magic angles\cite{Bistritzer12233}, has rapidly grown \cite{Mogera2020}.
Remarkable effects are seen in the optical properties by twisting bilayer graphene, for example, van Hove singularities lead to an energy band gap  lying in the visible spectrum of electromagnetic radiation\cite{Mogera2020}. Such gap is absent in any other known form of graphene\cite{Mogera2020}.
Although many studies are devoted to understand how flat bands arise in Dirac systems \cite{Tarnopolsky2019} and how they produce diverse quantum phases \cite{Yuan2018,ThreeBand}, still the effects under electromagnetic radiation are not well understood. Happily, a minimal model that also presents flat bands coexisting with Dirac cone states is the $\alpha-\mathcal{T}_{3}$ model, which consists of a honeycomb lattice with an additional atom located at the center of each hexagon and coupled with the atoms of just one of the two nonequivalent sublattices\cite{Sutherland1986}. While in twisted bilayer graphene the flat band arises at magic angles due to the multiple band crossings from the folding of the Brillouin zone caused by the moiré pattern\cite{Bistritzer12233,Tarnopolsky2019}; in the $\alpha-\mathcal{T}_{3}$ model the flat band comes from the local topology in the lattice\cite{Sutherland1986}. 

There are several achievable experimental systems that can be mapped to a low-energy $\alpha-\mathcal{T}_{3}$ model. These include the $\text{Hg}_{1-x}\text{Cd}_x$ quantum well\cite{HgCd-Malcom}, trilayer of cubic lattices\cite{Cubic-Wang} and optical lattices\cite{opticallattices01}. The $\alpha-\mathcal{T}_{3}$ model takes its name from the parameter $\alpha$, which stands for the coupling between sublattices, this parameter changes continuously the system from one with just two sublattices interacting (graphene plus a disconnected site) to one with the three sublattices equally coupled (dice lattice). Surprisingly enough, the dispersion relation for this system is $\alpha$ independent\cite{Vidal2001}. Nevertheless, breaking the symmetries in the system makes the physical quantities $\alpha$-dependent: introducing a potential barrier (breaking the spatial symmetry) results in an $\alpha$-dependent transmission\cite{Klein-nicol}; introducing a perpendicular magnetic field (breaking the time-reversal symmetry) makes the magneto-optical conductivity and the Hofstadter butterfly $\alpha$-dependent\cite{Magneto-Nicol,Magneto-Biswas,Magneto-kovacs,Biswas_2018}. In general, for the $\alpha-\mathcal{T}_{3}$ model, the flat band and the value of $\alpha$ become relevant in the presence of electromagnetic fields.  For example, the orbital susceptibility\cite{susceptibility-prl,piechon2015tunable} and the topological berry phase\cite{orbital-prb} can be tuned with the parameter $\alpha$,  and in contrast with graphene the plasmon branch is pinched in to a single point\cite{malcolm2016frequency}. More recently, it has been found that this kind of Hamiltonian leads to an unexpected family of in-gap chiral edge states for noninverted spin-1 Dirac quantum dots \cite{PhysRevResearch_Spin}. There are some  other previous works concerning the structure of the electromagnetic-dressed electron spectrum \cite{dey2018photoinduced,dey2019floquet,iurov2019peculiar}. These results indicate the opening of Floquet-band gaps, as has been determined before for other 2D materials as graphene \cite{Lopez_2008,RodriguezPhil,Kivis_2010,Foa_2011} and borophene \cite{Champo,Ibarra_2019}. This is essential to calculate photocurrents \cite{Lopez_2008} which is the usual  way to experimentally test the dynamic Floquet-bands  \cite{higuchi2017light,Heide_2018} and other optical properties \cite{Oliva_Leyva_2015,Oliva_2016}. For Dirac Hamiltonians with a flat-band, a study of the transitions induced by the electromagnetic fields is not available. In this work we present such study. Specially, we show how the inclusion of flat-band states shares many similarities with the three-level Rabi problem \cite{ThreeRabi}, in which the flat-band provides an intermediate step to mediate transitions from the valence to the conduction band. Our results show many useful relationships between transition probabilities, times and parameters of the model. This can serve to combine experimental and theoretical results to fine tune the parameters of the Hamiltonian, which is a problem that is still a work in progress \cite{Guinea2019}, as well as a tool to have quantum control of the system.

The layout of this work is the following. 
In Sec. \ref{sec:Model} we present the model and basic equations while in Sec. \ref{Sec:Electro} we include the electromagnetic field. Then in Sec.  \ref{Sec:Perturbation} we identify the transitions and relevant features of the model. Sec. \ref{Sec:Discussion} is devoted to present  the numerical results and a discussion is made to identify the relevant features of the model. Finally, in Sec. \ref{Sec:Conclusions} the conclusions are given.

\section{Hamiltonian for the $\alpha-\mathcal{T}_3$ model}\label{sec:Model}

In the tight-binding approximation, the $\alpha-\mathcal{T}_3$ Hamiltonian model considering only nearst-neighbor hopping is given by \cite{raoux2014dia,piechon2015tunable,malcolm2016frequency,dey2018photoinduced,iurov2019peculiar,dey2019floquet}
\begin{equation}\label{ec:Hamiltonian_Hk}
\hat{H}=\begin{pmatrix}
0 & t\mathcal{C}_\alpha\, f(\boldsymbol{k})& 0\\
t \mathcal{C}_\alpha\, f^{*}(\boldsymbol{k}) & 0 & t \mathcal{S}_\alpha\,  f(\boldsymbol{k})\\
0 & t \mathcal{S}_\alpha\, f^*(\boldsymbol{k}) & 0
\end{pmatrix},
\end{equation}
where $f(\boldsymbol{k})=\sum_{l=1}^3\mathrm{e}^{-i\boldsymbol{k}\cdot\boldsymbol{\delta}_l}$ and $t$ is the nearest-neighbor hopping amplitude, with $\boldsymbol{\delta}_i$ ($i=1, 2, 3$) the vectors connecting the nearest-neighbors sites and $\boldsymbol{k}=(k_x,k_y)$ the momentum vector. The low-energy Hamiltonian around the two inequivalent Dirac points can be written as\cite{malcolm2016frequency,dey2018photoinduced}
\begin{equation}\label{ec:LowEnergy_H}
\hat{H}^{0}_{\xi}=\hbar v_F\boldsymbol{k}\cdot\boldsymbol{\hat S}, 
\end{equation}
where  $v_F\approx 10^{6}$ m/s is the Fermi velocity and $\boldsymbol{\hat S}=(\xi \hat{S}_x,\hat{S}_y)$ with $\xi =\pm 1$ refers to the \textbf{K} and  \textbf{K}' valleys. The pseudo-spin operators $\hat{S}_x$ and $\hat{S}_y$ are defined as
\begin{eqnarray}
\hat{S}_x&=&
\begin{pmatrix}\label{ec:SpinOpe_Sx}
0 & \mathcal{C}_\alpha & 0\\ 
\mathcal{C}_\alpha & 0 & \mathcal{S}_\alpha\\
0 & \mathcal{S}_\alpha & 0 
\end{pmatrix},\\
\hat{S}_y&=&
\begin{pmatrix}\label{ec:SpinOpe_Sy}
0 & -i\mathcal{C}_\alpha & 0\\ 
i\mathcal{C}_\alpha & 0 & -i \mathcal{S}_\alpha\\
0 & i \mathcal{S}_\alpha & 0 
\end{pmatrix},
\end{eqnarray}
with $ \mathcal{C}_\alpha=1/\sqrt{1+\alpha^2}$, $ \mathcal{S}_\alpha=\alpha/\sqrt{1+\alpha^2}$ and $0\leq \alpha\leq 1$. From Eq. \eqref{ec:Hamiltonian_Hk} is possible to find the electronic band structure and the eigenfunctions by solving the eigenvalue problem 
\begin{equation}\label{ec:EigenvalueProblem}
\hat{H}^{0}_{\xi} \ket{\psi_{\boldsymbol{k},\mu}}=E_{\mu} \ket{\psi_{\boldsymbol{k},\mu}},   
\end{equation}
where the band structure consists in two Dirac cones and an additional flat band. The first cone is described by $E_{1}=- v_F\hbar|\boldsymbol{k}|$ that correspond to valence band (VB), the flat band (FB) is described by $E_{2}=0$, and the second cone is $E_{3}=v_F\hbar|\boldsymbol{k}|$ and describes the conduction band (CB). In the Fig. (\ref{fig:BandsTransitions}), we show this band structure for near the Dirac point \textbf{K} ($\xi=+1$).

\begin{figure}[t]
\includegraphics[scale=0.6]{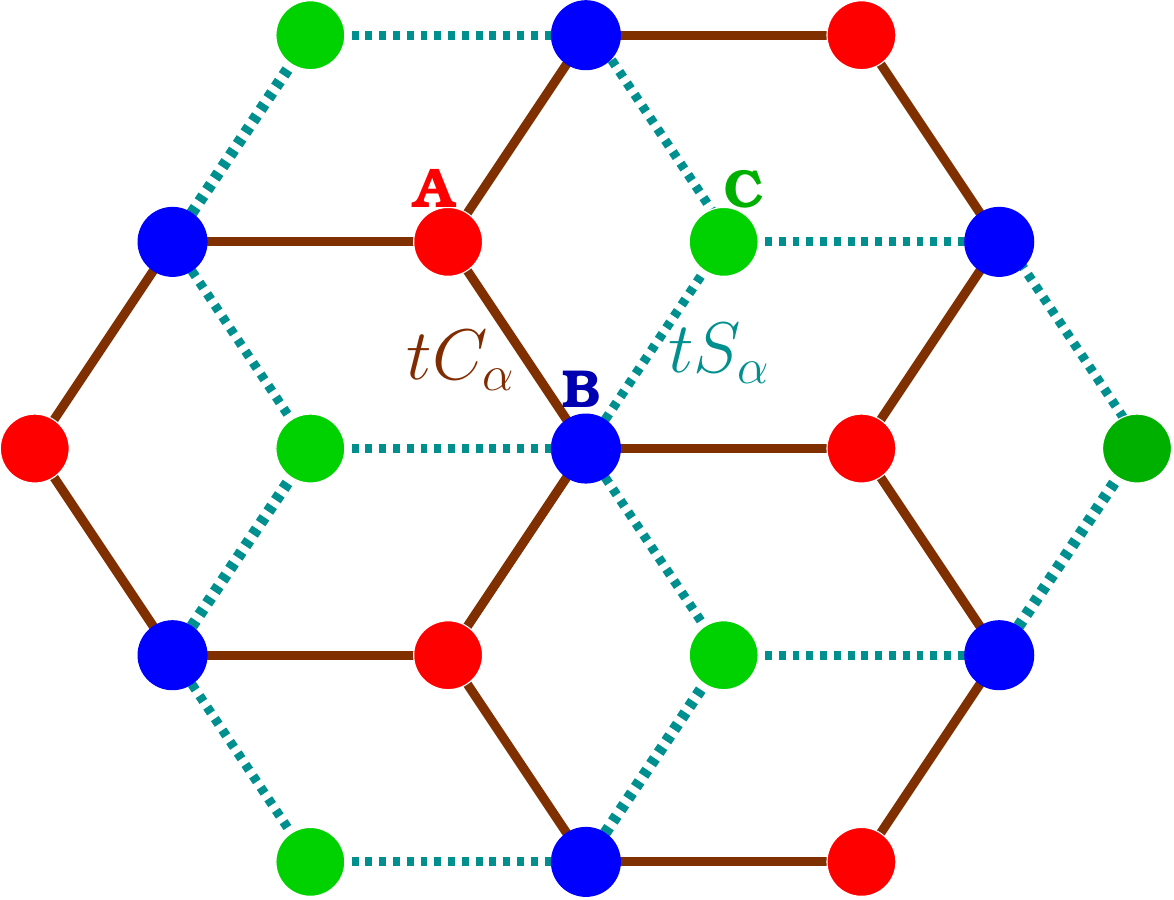}
\caption{\label{fig:Lattice} Sketch of the $\alpha- \mathcal{T}_3$ lattice. When $\alpha=0$ ($\mathcal{C}_\alpha=1$, $\mathcal{S}_\alpha=0$) results in the honeycomb lattice resembling monolayer graphene. In contrary, for $\alpha=1$  ($\mathcal{C}_\alpha=\mathcal{S}_\alpha=1/\sqrt{2}$) leads to the well-studied  dice lattice with pseudospin 1 \cite{malcolm2016frequency}.}
\end{figure}

The eigenfunctions for each band are given by 
\begin{eqnarray}
\ket{\psi_{\boldsymbol{k},1}}&=&\frac{1}{\sqrt{2}}\Big[\xi \mathcal{C}_\alpha\mathrm{e}^{-i\xi\theta_{\boldsymbol{k}}}\ket{\mathrm{A}}-\ket{\mathrm{B}}+\xi \mathcal{S}_\alpha\mathrm{e}^{i\xi\theta_{\boldsymbol{k}}}\ket{\mathrm{C}}\Big],\\
\ket{\psi_{\boldsymbol{k},2}}&=&\xi \mathcal{S}_\alpha\mathrm{e}^{-i\xi\theta_{\boldsymbol{k}}}\ket{\mathrm{A}}-\xi \mathcal{C}_\alpha\mathrm{e}^{i\xi\theta_{\boldsymbol{k}}}\ket{\mathrm{C}},\\
\ket{\psi_{\boldsymbol{k},3}}&=&\frac{1}{\sqrt{2}}\Big[\xi \mathcal{C}_\alpha\mathrm{e}^{-i\xi\theta_{\boldsymbol{k}}}\ket{\mathrm{A}}+\ket{\mathrm{B}}+\xi \mathcal{S}_\alpha\mathrm{e}^{i\xi\theta_{\boldsymbol{k}}}\ket{\mathrm{C}}\Big],
\end{eqnarray}
where $\theta_{\boldsymbol{k}}=tan^{-1}(k_y/k_x)$, and 
\begin{equation}
\ket{\mathrm{A}}=
\begin{pmatrix}
1\\
0\\
0
\end{pmatrix},\,\,\,\,\,
\ket{\mathrm{B}}=
\begin{pmatrix}
0\\
1\\
0
\end{pmatrix},\,\,\,\,\,
\ket{\mathrm{C}}=
\begin{pmatrix}
0\\
0\\
1
\end{pmatrix},
\end{equation}
are the spinors that describe the sublattice degree of freedom ($\mathrm{A}$, $\mathrm{B}$ and $\mathrm{C}$) in the unit cell as shown in Fig. \ref{fig:Lattice}.

\section{The $\alpha-\mathcal{T}_3$ model under electromagnetic radiation and interaction picture}\label{Sec:Electro}

To study the dynamics of the electrons in the $\alpha-\mathcal{T}_3$ model under electromagnetic radiation, we introduce a minimal coupling through the Peierls substitution\cite{dey2018photoinduced} $\hbar\boldsymbol{k}\rightarrow\hbar\boldsymbol{k}-e\boldsymbol{A}$ in the low-energy Hamiltonian \eqref{ec:LowEnergy_H}, where $\boldsymbol{A}=(A_x,A_y)$ is the vector potential of the electromagnetic wave. For this problem, we consider a gauge in which the components $A_x$ and $A_y$ are only a function of time. 
 
 From the Eq. \eqref{ec:LowEnergy_H} we obtain:
\begin{equation}\label{ec:DiracHalmitonianElectromagneticField} 
\hat{H}_{\xi}(t)=\hat{H}^{0}_{\xi}+\hat{V}_{\xi}(t), 
\end{equation}
where $\hat{H}^{0}_{\xi}$ is given by Eq. \eqref{ec:LowEnergy_H} and $\hat{V}_{\xi}(t)$ is defined as
\begin{equation}\label{ec:TimePotential}
\hat{V}_{\xi}(t)=-e v_F\boldsymbol{A}\cdot\boldsymbol{\hat S}.   
\end{equation}
In the Eqs. \eqref{ec:LowEnergy_H} and $\eqref{ec:TimePotential}$ the vectors $\hbar v_F\boldsymbol{k}$ and  $e v_F\boldsymbol{A}$ represent the directional energy flux of electrons and the work done by the electromagnetic wave along the $x$ and $y$ directions, respectively.

In the interaction picture \cite{sakurai1995modern}, the Dirac equation for the $\alpha-\mathcal{T}_3$ model is given by
\begin{equation}\label{ec:DiracEquationInteraction}
i\hbar\frac{d}{dt}\boldsymbol{\chi}(t)=\mathbbm{V}(t)\boldsymbol{\chi}(t),
\end{equation}
where the vector $\boldsymbol{\chi}(t)= \Big(\chi_{1}(t),\,\chi_{2}(t),\,\chi_{3}(t)\Big)^{\top}$ contains the components of the wavefunctions in the VB ($\mu=1$), FB ($\mu=2$), and CB ($\mu=3$) in the interaction picture and they are given by
\begin{eqnarray}\label{ec:VectorComponents}
\chi_{\mu}(t)&=&\exp\left[i \frac{E_\mu}{\hbar}t\right]\braket{\psi_{\boldsymbol{k},\mu}|\boldsymbol{\Psi}(t)},
\end{eqnarray}
and $\ket{\boldsymbol{\Psi}(t)}$ is a time-dependent three-component spinor in the Schr\"oendiger picture. In Eq. \eqref{ec:DiracEquationInteraction}, $\mathbbm{V}(t)$ is a square matrix with dimensions $3\times3$ and its components are defined as
\begin{equation}\label{ec:MatrixVInteraction}
[\mathbbm{V}(t)]_{\mu,\nu}=\exp\left[i \frac{(E_\mu-E_\nu)}{\hbar}t\right]\bra{\psi_{\boldsymbol{k},\mu}}\hat{V}_{\xi}(t)\ket{\psi_{\boldsymbol{k},\nu}},    
\end{equation}
where the subindeces $\mu,\, \nu=1,\,2,\,3$ refer to the band.

Let now study the case of a linearly polarized electromagnetic wave defined by the vector potential
\begin{equation}\label{ec:VectorPotential}
\boldsymbol{A}=\frac{E_0}{\Omega}\cos(\Omega t)\boldsymbol{\hat{r}},
\end{equation}
where $\boldsymbol{\hat{r}}=\left(\cos\Theta,\,\sin\Theta\right)$ is the polarization vector, $E_0$ is the amplitude of the electric field, taken as constant, and $\Omega$ is the frequency of the electromagnetic wave. Notice that here we consider a classical field, and thus we are assuming a quantum coherent field with a huge number of photons\cite{shirley1965solution}.   Without any loss of generality, we can take $\Theta=0$ as the physics
only depends upon the  angle between $\Theta$ and
$\theta_{\boldsymbol{k}}$.

Using the Eq. \eqref{ec:VectorPotential} and rewriting the Eqs. \eqref{ec:DiracEquationInteraction}-\eqref{ec:MatrixVInteraction}, we obtain
\begin{equation}\label{ec:DiracEquationLiearlyPolarized}
i\boldsymbol{\chi}'(t)+\cos(\Omega t)\mathbbm{B}(t)\ \boldsymbol{\chi}(t)=0,
\end{equation}
where $\mathbbm{B}(t)$ is a matrix defined as
\begin{equation}\label{ec:MatrixBLinearlyPolarized}
\mathbbm{B}(t)= 
\begin{pmatrix} 
-\epsilon & w\,\mathrm{e}^{-i \omega t} & s\,\mathrm{e}^{-i2\omega t}\\ 
w^{*}\,\mathrm{e}^{i \omega t} & 0 & w \mathrm{e}^{-i \omega t}\\
s^{*}\,\mathrm{e}^{2 i \omega t}  & w^{*} \mathrm{e}^{ i \omega t} & \epsilon\\
\end{pmatrix}.
\end{equation}
The coefficients $\epsilon$, $w$ and $s$ from the previous expression are defined as
\begin{equation}\label{ec:epsilon_coeficient}
\epsilon=\zeta\cos \theta_{\boldsymbol{k}},
\end{equation}
\begin{equation}\label{ec:w_coeficient}
w=i\sqrt{2}\xi\mathcal{C}_\alpha\mathcal{S}_\alpha\zeta\sin \theta_{\boldsymbol{k}},    
\end{equation}
\begin{equation}\label{ec:s_coeficient}
s=i\xi(\mathcal{C}^2_\alpha-\mathcal{S}^2_\alpha)\zeta\sin \theta_{\boldsymbol{k}} ,
\end{equation}
and the coefficient
\begin{equation}\label{ec:ZetaParameter}
\zeta=\frac{e v_{F} E_0}{\hbar\Omega}.    
\end{equation}
In Eq. \eqref{ec:MatrixBLinearlyPolarized} we used a set of renormalized moments, $\kappa_x=(v_F/\Omega)k_x$ and $\kappa_y=(v_F/\Omega)k_y$, which are related with $\omega$ as follows:
$\omega=\Omega\sqrt{\kappa_x^2+\kappa_y^2}$. 

\section{Transitions}\label{Sec:Perturbation}

This section is devoted to study the transitions that are obtained from Eq. (\ref{ec:DiracEquationLiearlyPolarized}). The first observation concerning Eq. (\ref{ec:DiracEquationLiearlyPolarized}) is that the system depends upon the angle $\theta_{\boldsymbol{k}}$. For some angles, the solutions of Eq. (\ref{ec:DiracEquationLiearlyPolarized})  are readily found by direct integration, while others require perturbation theory, as detailed below.

\subsection{Parallel or antiparallel incidence angle}

When the electron momentum is parallel  to the electric field, i.e., for $\theta_{\boldsymbol{k}}=0$ (also for the anti-parallel $\theta_{\boldsymbol{k}}=\pi$ the solution is similar), we have that $w=0$ and $s=0$. Therefore, we have in principle that for $\eta=1,3$,
\begin{equation}
\chi_{\eta}(t)=\chi_{\eta}(0)\exp\left[\mp i\frac{\zeta}{\Omega}\sin {(\Omega t)}\right],
\end{equation}
where the upper sign is for $\eta=1$ and the lower for $\eta=3$. We also have $\chi_{2}(t)=\chi_{2}(0)$. Using
Eq. \eqref{ec:VectorComponents} we obtain for $\eta=1,3$,
\begin{equation}\label{eq:oneandtree}
\braket{\psi_{\boldsymbol{k},\eta}|\boldsymbol{\Psi}(t)}= \chi_{\eta}(0)\exp\left[\mp \frac{i\zeta}{\Omega} \sin{\Omega t}\right]\exp\left[-i \frac{E_{\eta}}{\hbar}t\right],
\end{equation}
while for the flat-band,
\begin{equation}\label{eq:flatparallel}
\braket{\psi_{\boldsymbol{k},2}|\boldsymbol{\Psi}(t)}= \chi_{2}(0)\exp\left[-i \frac{E_{2}}{\hbar}t\right].
\end{equation}
Eqs. \eqref{eq:oneandtree} and \eqref{eq:flatparallel} show that band  occupation probabilities are constant over time.

It is interesting to use Floquet theory to analyze such result. We decompose Eq. \eqref{eq:oneandtree} as a Fourier series,
\begin{multline}
\braket{\psi_{\boldsymbol{k},\eta}|\boldsymbol{\Psi}(t)}=\chi_{\eta}(0)\\ \times\sum_{m=1}^{\infty}(-i)^{m}J_{m}\left(\frac{\zeta}{\Omega}\right)\exp\left[-\frac{i}{\hbar}\left(E_\eta+m\hbar\Omega\right)t\right],
\end{multline}
where $J_{m}(x)$ denotes the $m$-esim Bessel function. The argument of the exponential can be identified with the Floquet quasienergy $\epsilon_m=E_{\eta}+m\hbar \Omega$, with a first-Brillouin zone determined by $\hbar \Omega$.  Here $|J_{m}(\zeta/\Omega)|^{2}$ represents the amplitude of emission or absorption for $m=1,2,3,4,...$ photon processes starting from a coherent field, i.e., a field in which there is a huge number of photons present \cite{shirley1965solution}. It is surprising  to have photon emission and absorption while keeping
the occupation probabilities the same. However, this is easy to understand if we observe that after a time
period given by $T=2\pi/\Omega$, the wave-function only gets a phase given by,

\begin{equation}\label{eq:phase}
\braket{\psi_{\boldsymbol{k},\eta}|\boldsymbol{\Psi}(t+T)}=\exp{\left[\frac{2i\pi}{\hbar\Omega}\right]}\braket{\psi_{\boldsymbol{k},\eta}|\boldsymbol{\Psi}(t)}.
\end{equation}
 When the field is at resonance with a given $E_{\mu}$, i.e., say for example $E_{1}=n\hbar\Omega$ with $n$ an integer, the change of phase in eq. \eqref{eq:phase} is $2\pi n$ for $\eta=1$ and $-2\pi n$ for $\eta=3$. All these results are explained from the fact that the field does not break the symmetry along the $x$-axis  for this incidence.
 As we will see below, this is not the case for other incidence angles.

\subsection{General incidence angle: Time-dependent perturbation theory}\label{Subsection:perturbation}

Obtaining an analytical solution for the system of equations that arise from the Eq. \eqref{ec:DiracEquationLiearlyPolarized} for $\theta_{\boldsymbol{k}} \neq 0$ or $\theta_{\boldsymbol{k}} \neq \pi$ is difficult.  Nevertheless, it is possible to find information on the dynamics of $\alpha-\mathcal{T}_3$ in the interaction picture using time-dependent perturbation theory \cite{sakurai1995modern}.

In this approach, the solution for each component of the Eq.\eqref{ec:DiracEquationLiearlyPolarized} can be expressed as follows,
\begin{equation}\label{ec:InteractionExpansion}
\chi_{\mu}(t)= \chi^{(0)}_{\mu}(t)+ \chi_{\mu}^{(1)}(t) + ...\,,   
\end{equation}
where  $\chi_{\mu}^{(0)}(t)$ and  $\chi_{\mu}^{(1)}(t)$ signify amplitudes of zero order, first order, and so on, in the strength parameter $\zeta$ of the time-dependent potential. 

To zero order the amplitudes are time independent, 
\begin{equation}\label{ec:AmplitudeZeroOrder}
\chi_{\mu}^{(0)}=\sum_{\nu}\braket{\psi_{\boldsymbol{k},\mu}|\psi_{\boldsymbol{k},\nu}}=\sum_{\nu} \delta_{\mu,\nu},     
\end{equation}
where $\ket{\psi_{\boldsymbol{k},\nu}}$ denotes the initial state $\nu$. The first order correction is,
\begin{equation}\label{ec:AmplitudeFirstOrder}
\chi_{\mu}^{(1)}(t)=\sum_{\nu}a_{\mu\nu}(t),
\end{equation}
where $a_{\mu\nu}(t)$ is given by,
\begin{eqnarray}\label{ec:Amunu}
a_{\mu \nu}(t)&=&-|B_{\mu \nu}| \int_{0}^{t}   \mathrm{e}^{i \arg[B_{\mu \nu}(t')]} \cos(\Omega t')dt',
\end{eqnarray}
with $B_{\mu,\nu}$ are the coefficients defined in the Eq. \eqref{ec:MatrixBLinearlyPolarized}. For the transitions from the VB ($\mu=1$) to the FB ($\nu=2$) we obtain,
\begin{eqnarray}\label{ec:AmplitudeFirstOrder}
a_{21}(t)
&=&-\frac{i w}{2}\left\{\frac{1-\exp\Big[i(\omega+\Omega )t\Big]}{\omega+\Omega}\right\}\nonumber\\
&&-\frac{i w}{2}\left\{\frac{1-\exp\Big[i(\omega-\Omega)t\Big]}{\omega-\Omega}\right\},
\end{eqnarray}
The last expression is formed by two terms. The first corresponds to a photon emission process while the second represents an absorption. 

The squared amplitudes of the coefficients $a_{\mu\nu}(t)$ give the transition probability from the state $\nu$ to the state $\mu$.  At a given frequency of the electromagnetic wave, $\Omega$, and consider the photon absorption mechanism there will be two resonant frequencies. Namely, (1) $\omega \approx \Omega$ which is associated with two simultaneously transitions from VB to FB and from FB to CB; and (2) $\omega \approx \Omega/2$ related with the transition from VB to CB.

\begin{figure}[t]
\includegraphics[scale=0.35]{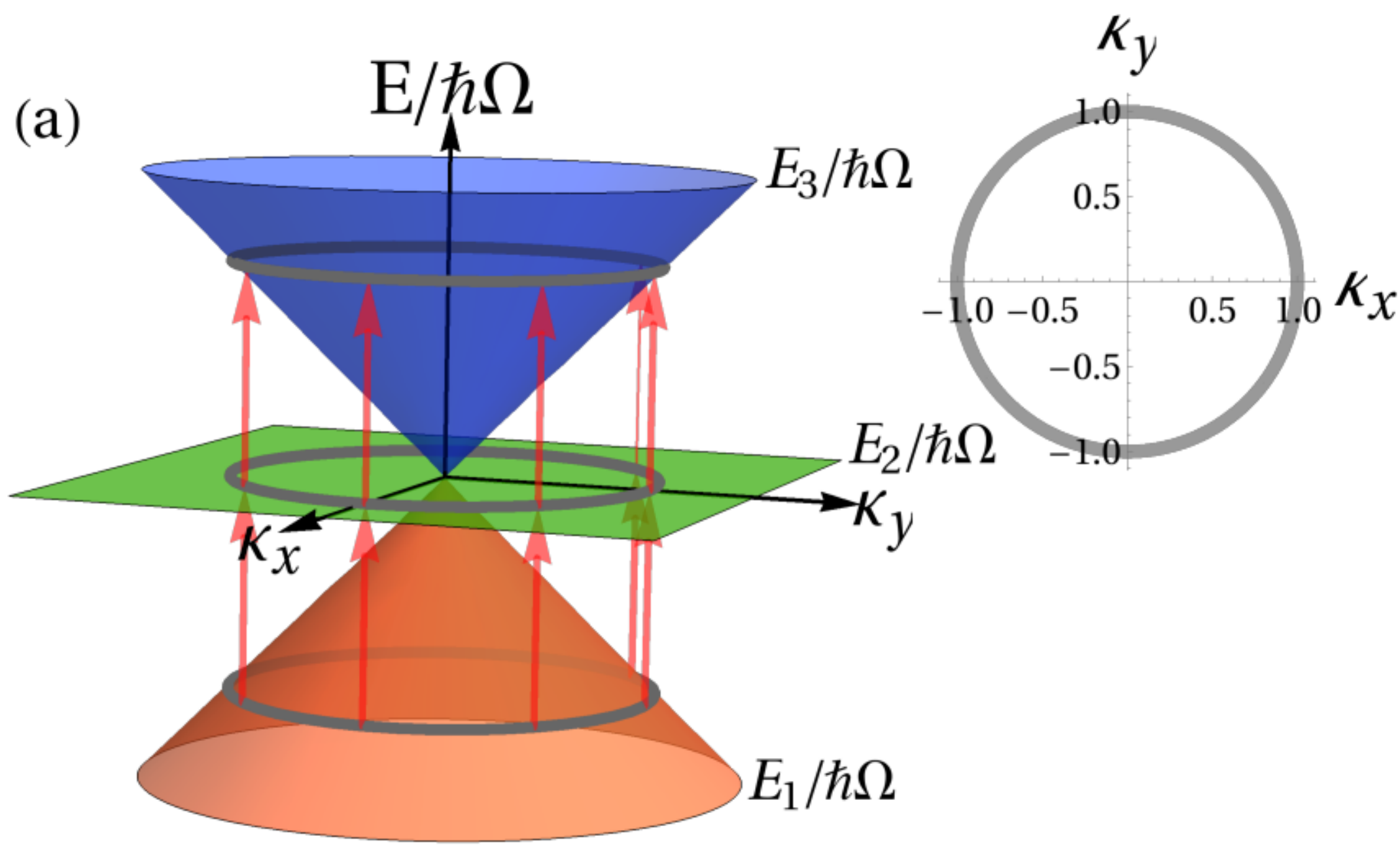}
\includegraphics[scale=0.35]{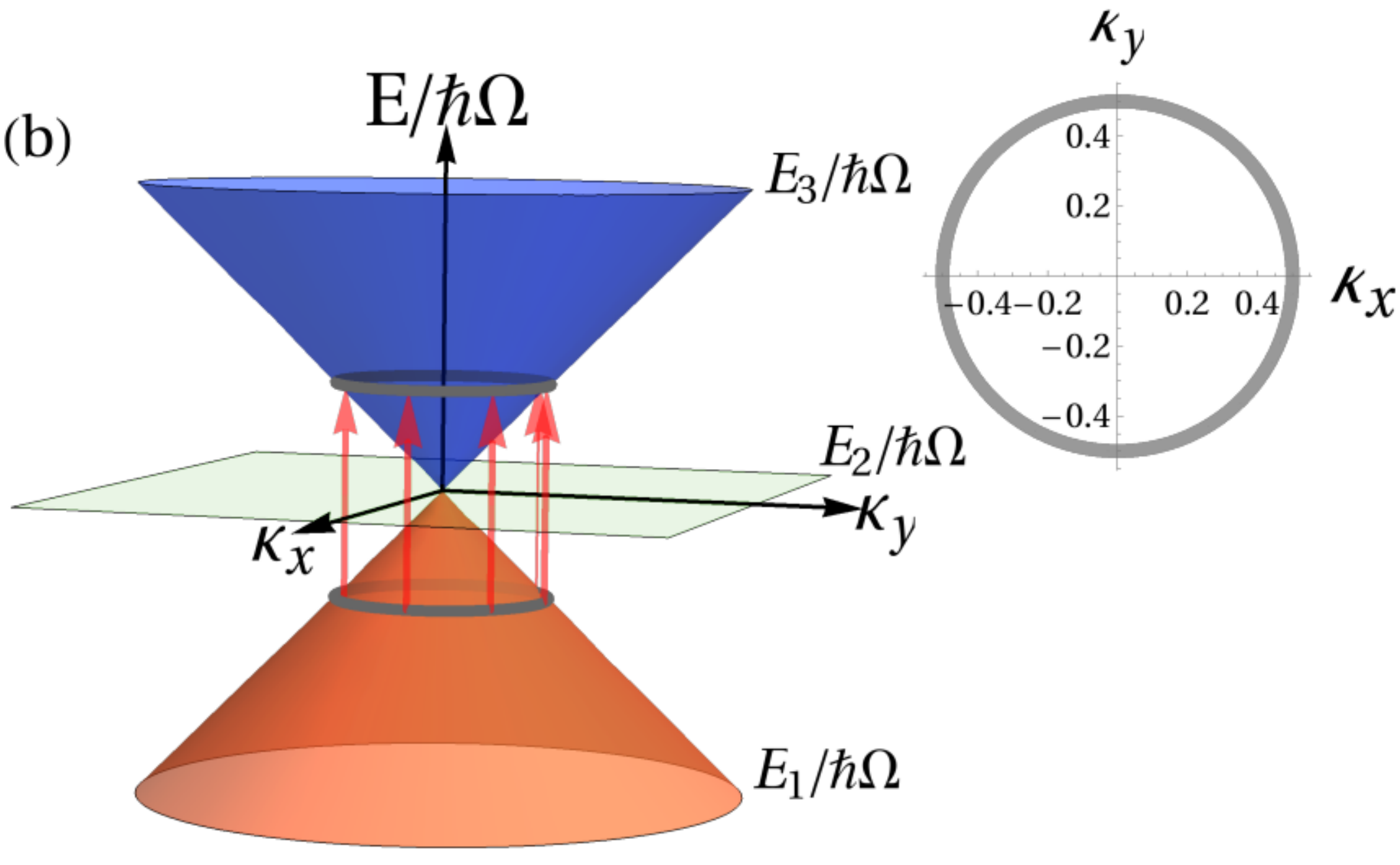}
\caption{\label{fig:BandsTransitions} Band structure of the $\alpha-\mathcal{T}_3$ lattice around the Dirac point \textbf{K}. The light-blue Dirac cone correspond to the CB ($E_{3}$), light-green is the FB ($E_{0}$) and light-orange is the VB ($E_{1}$). When an electromagnetic field with a given constant frequency $\Omega$ is introduced, direct interband transitions denoted by light-red arrows are allowed. In (a) shows the transitions for the resonant frequency $\omega\approx\Omega$ (corresponding to $\kappa_x^2+\kappa_y^2 \approx 1$ as indicated by the grey circle) 
which is associated with two simultaneously transitions from the VB to the FB and from the FB to the CB.  In (b), we show the transition for the resonant frequency $\omega\approx\Omega/2$ (corresponding to $\kappa_x^2+\kappa_y^2\approx1/4$ as indicated by the grey circle) which is associated with the transitions from the VB to the CB .
}
\end{figure}

Therefore, using Eq. \eqref{ec:Amunu} and omitting the photon emission mechanism, we obtain
\begin{equation}
\mathcal{P}^{A}_{2\leftarrow 1}=|w|^2 \frac{\sin^2\left[\left(\omega-\Omega\right)t/2\right]}{\left(\omega-\Omega\right)^2} \, ,
\end{equation}
and,
\begin{equation}
 \mathcal{P}^{A}_{3\leftarrow 2}= \mathcal{P}^{A}_{2\leftarrow 1} \, ,
\end{equation}
for the former case, while for the transition that does not involve the flat band 
\begin{equation}
\mathcal{P}^{A}_{3\leftarrow 1}=|s|^2 \frac{\sin^2\left[\left(2 \omega-\Omega\right)t/2\right]}{\left(2\omega-\Omega\right)^2}\, .   
\end{equation}

Hence, for transition probabilities $ \mathcal{P}^{A}_{3\leftarrow 2}$ and $\mathcal{P}^{A}_{2\leftarrow 1}$ in the resonance frequency $\omega\approx \Omega$, and using the fact that $\omega=\Omega\sqrt{\kappa_x^2+\kappa_y^2}$, we find that the maximum transitions occurs when the renormalized moments satisfy  $\kappa_x^2+\kappa_y^2\approx1$. In the Fig. \ref{fig:BandsTransitions} (a), we show these transitions probabilities for this resonant frequency. Similarly, for transitions $ \mathcal{P}^{A}_{3\leftarrow 1}$ and $\omega\approx \Omega/2$, the maximum transition is given by $\kappa_x^2+\kappa_y^2\approx (1/2)^2$ as a show in the Fig \ref{fig:BandsTransitions} (b). 

\section{Exact numerical results and discussion}\label{Sec:Discussion}

In this section, we study the behavior of band occupation probabilities $|\chi_{\mu}(t)|^2$. To simplify the discussion, here we will focus in the particular case when the electron momentum is perpendicular to the electric field ($\theta_{\boldsymbol{k}}=\pi/2$) and in the \textbf{K} valley ($\xi=+1$).  The results presented here are representative for other values of $\theta_{\boldsymbol{k}}$ and for the \textbf{K'} valley ($\xi=-1$). Then we solve numerically the system of coupled differential equations that comes from Eq. \eqref{ec:DiracEquationLiearlyPolarized}. For this purpose, we consider that the amplitude of the electric field has a value $E_0=5\times 10^{-2}\,$V/m and the frequency of the electromagnetic wave is $\Omega=22\,$GHz. With these values and using Eq. \eqref{ec:ZetaParameter}, the strength parameter is  $\zeta=3.45\,$GHz. In the following, we will discuss the occupation probabilities in the cases of two resonant frequencies discussed in section \ref{Sec:Perturbation}.

\subsection{Resonance frequency $\omega\approx\Omega$: Three and two-level Rabi problems}

\begin{figure}[t]
\begin{center}
\includegraphics[scale=0.65]{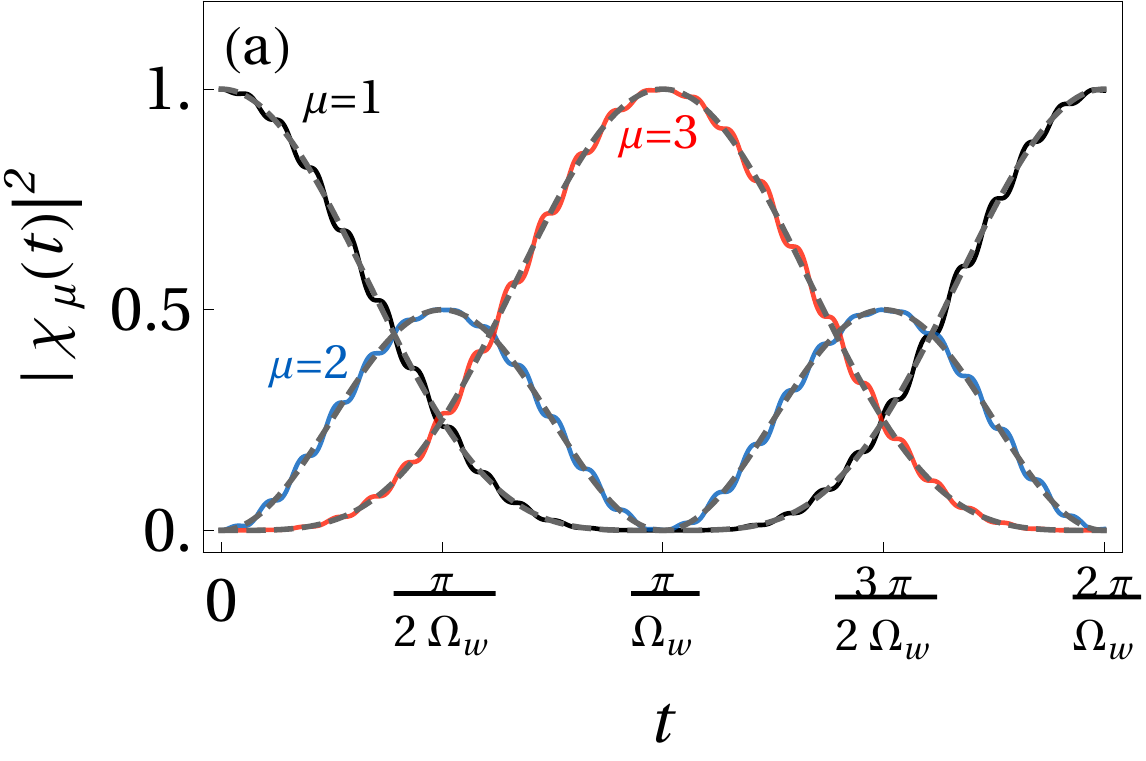}
\includegraphics[scale=0.65]{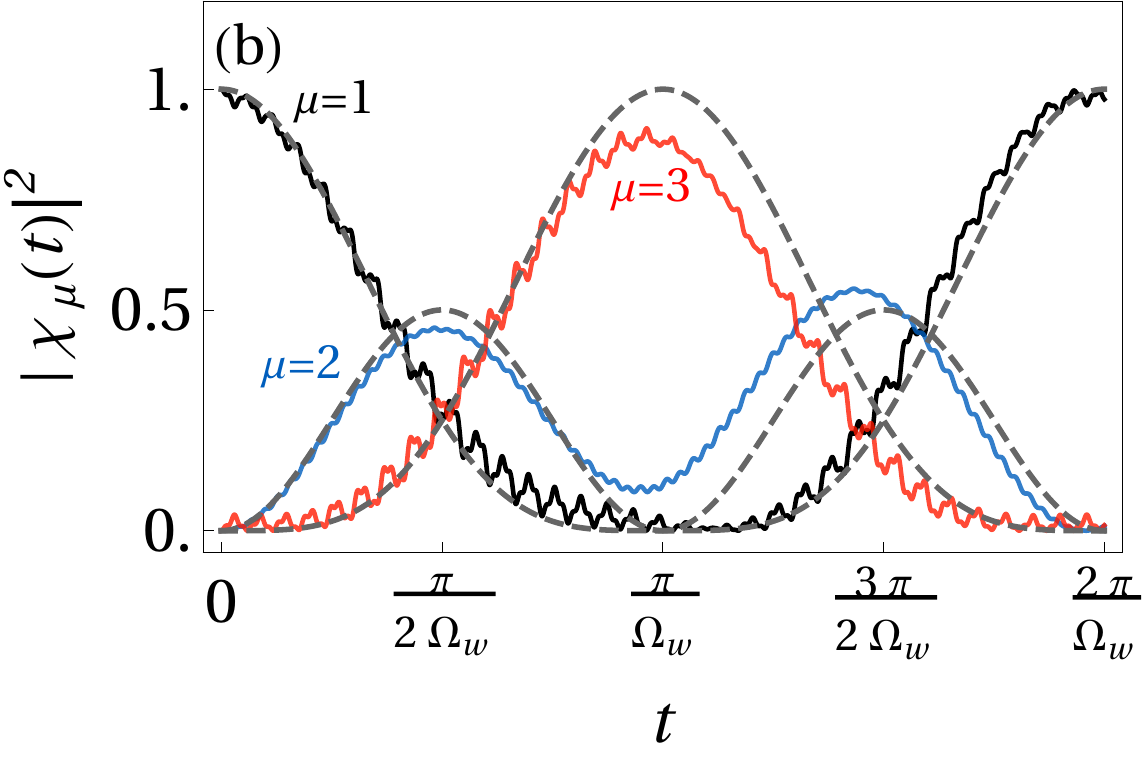}
\end{center}
\caption{\label{Fig:ProbabilitiesThreeRabiA}Occupation probabilities $|\chi_{1}(t)|^2$ (black line),  $|\chi_{2}(t)|^2$ (red line) and   $|\chi_{3}(t)|^2$ (blue line) as a function of time for the resonant frequency $\omega\approx\Omega$. The dotted gray lines are the results obtained for the three-level Rabi problem considering only resonant terms. In, (a) shows the numerical results for $\alpha=1.0$ and $\Omega_w=1.72$ Gz. In, (b) shows the behavior for $\alpha=0.2$ with $\Omega_w=0.66$ Gz.  The initial state in both panels are given by $\chi_{1}(0)=1$. Note that although $\Omega_{w}$ is different in both panels, the periodicity is determined by $\Omega_{w}$. Notice how the case $\alpha=1$ corresponds to the three-level Rabi problem.}
\end{figure}

As we saw in subsection \ref{Subsection:perturbation}, one resonant frequency is at $\omega\approx\Omega$. We  consider $\theta_{\boldsymbol{k}}=\pi/2$, for which the normalized moments take the values $\kappa_x\approx0$ and $\kappa_y \approx 1 $.

In the Fig. \ref{Fig:ProbabilitiesThreeRabiA}(a) we plot the occupation probabilities $|\chi_{\mu}(t)|^2$ for the VB (black line), FB (blue line) and CB (red line) for $\alpha=1$, when the initial state is $\chi_1(0)=1$ (numerical solution). For this value of $\alpha$, the bond between sites A and B, B and C are the same (see Fig. \ref{fig:Lattice}). Accordingly, $\mathcal{C}_\alpha=\mathcal{S}_\alpha=1/\sqrt{2}$. This reduces the coupling of the system of differential equations in Eq. \eqref{ec:DiracEquationLiearlyPolarized} as $s=0$. Therefore, it turns out that the matrix $\mathbbm{B}(t)$ 
is the same of a know simplified solvable version of the three-level Rabi system \cite{ThreeRabi,Shore_1997} and that we can can compare with our numerical results. The main difference between the three-level and two-level Rabi system is that it can be used to probe the medium through the interaction with a first transition. According to Sargent and Horwitz, the band-occupances are given by\cite{ThreeRabi},
\begin{eqnarray}\label{eq:ThreeRabi}
|\chi_{1}(t)|^2&=&\left[\cos\left(\frac{\Omega_w}{2} t\right)\right]^4,\label{eq:ThreeRabiOne}\\
|\chi_{2}(t)|^2&=&\frac{1}{2}\left[\sin\left(\Omega_w t\right)\right]^2,\label{eq:ThreeRabiTwo}\\
|\chi_{3}(t)|^2&=&\left[\sin\left(\frac{\Omega_w}{2} t\right)\right]^4\label{eq:ThreeRabiThree}.
\end{eqnarray}
The band-occupation probabilities present oscillations with period for the BV and BC transitions equal to $T=2\pi/\Omega_w$ where $\Omega_w=|w|/\sqrt{2}$ with $|w|$ defined by Eq. \eqref{ec:w_coeficient}. In this particular case we have  $\Omega_w=1.72\,$GHz. For the FB transitions, the oscillation period is $T/2$. The maximum amplitude of oscillations for VB and CB is equal to $1$. This is a consequence of the resonant frequency and $\alpha=1$. Notice that although the external field frequency is resonant for transitions between the FB to the cone levels, the flat state can only be half-occupied. The reason is that the probability leaks from the valence band to the conduction band through the intermediate FB\cite{ThreeRabi}. This can be corroborated by observing that for this case, time perturbation theory indicates that $\mathcal{P}^{A}_{3\leftarrow 1}=0$ while $\mathcal{P}^{A}_{2\leftarrow 1}=\mathcal{P}^{A}_{3\leftarrow 2}$.

\begin{figure}[t]
\begin{center}
\includegraphics[scale=0.65]{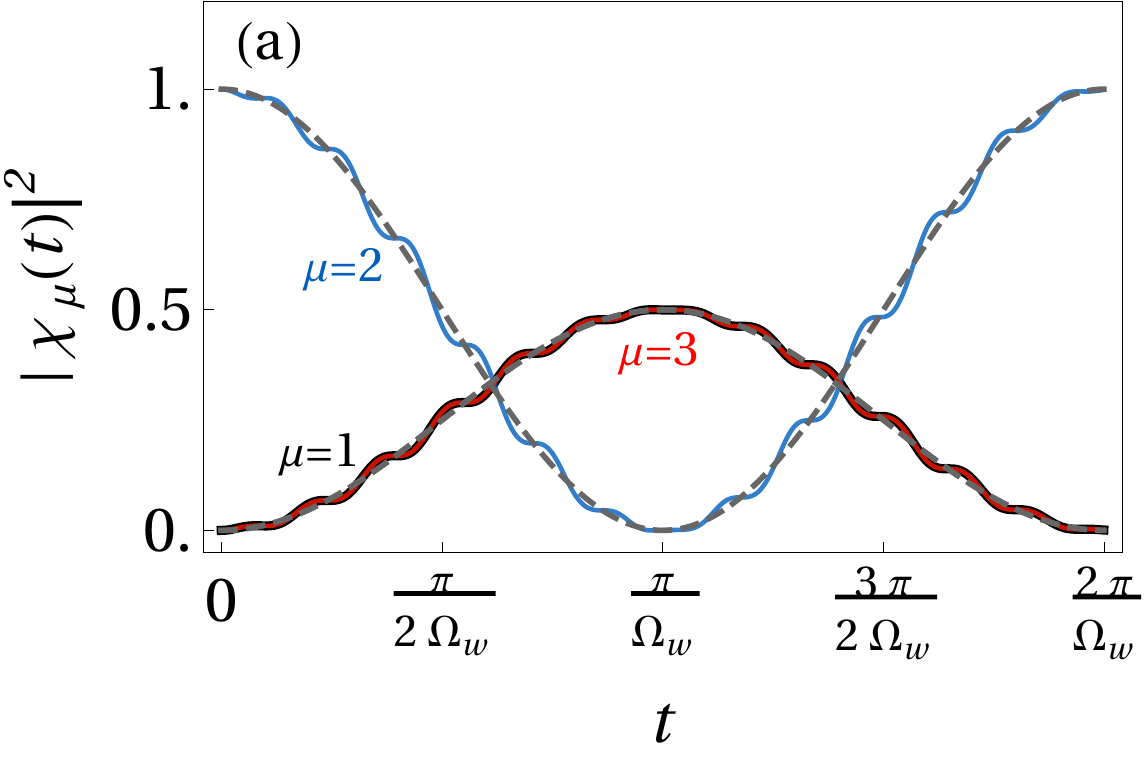}
\includegraphics[scale=0.65]{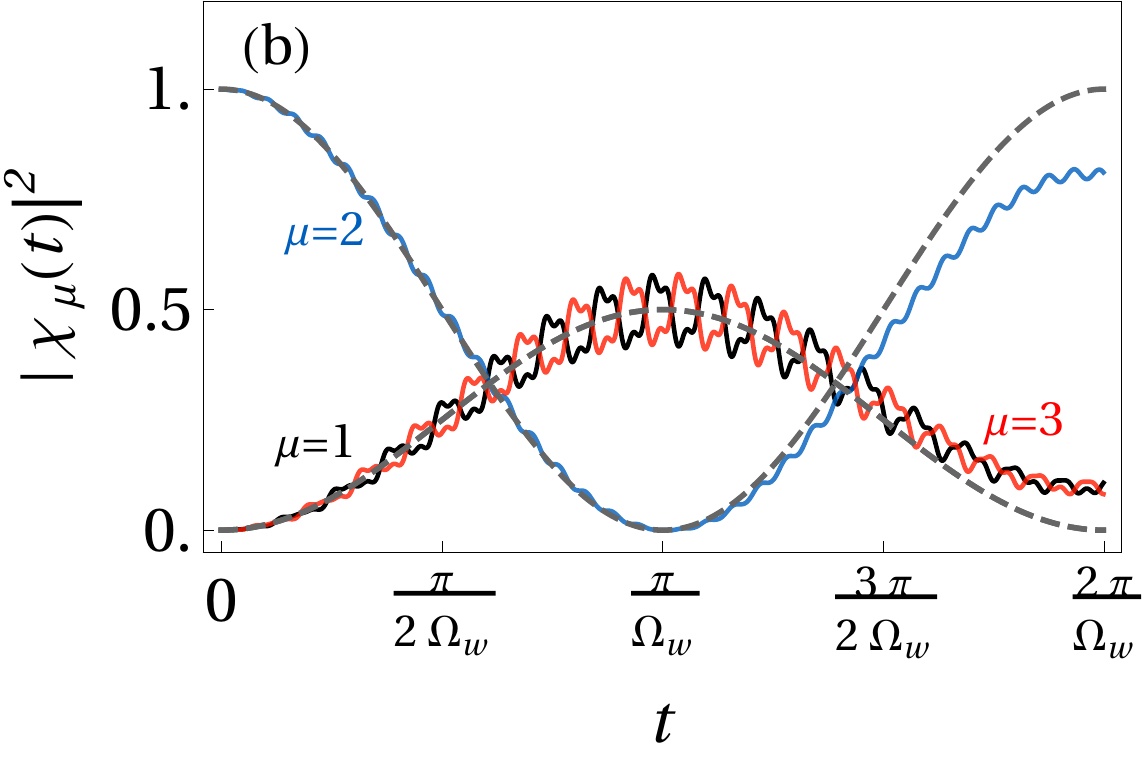}
\end{center}
\caption{\label{Fig:ProbabilitiesThreeRabiB}Occupation probabilities $|\chi_{1}(t)|^2$ (black line),  $|\chi_{2}(t)|^2$ (red line) and   $|\chi_{3}(t)|^2$ (blue line) as a function of time for the resonant frequency $\omega\approx\Omega$. The dotted gray lines are the results obtained for the three-level Rabi problem considering only resonant terms. In, (a) shows the numerical results for $\alpha=1.0$ and $\Omega_w=3.45$ Gz. In, (b) shows the behavior for $\alpha=0.2$ with $\Omega_w=1.32$ Gz.  The initial state in both panels are given by $\chi_{2}(0)=1$. Note that although $\Omega_{w}$ is different in both panels, the periodicity is determined by $\Omega_{w}$. Notice how the case $\alpha=1$ corresponds to the three-level Rabi problem}
\end{figure}

In Fig. \ref{Fig:ProbabilitiesThreeRabiA}(a) we show a comparison between the band occupancies given by Eqs. \eqref{eq:ThreeRabiOne}-\eqref{eq:ThreeRabiThree} (gray dotted lines) with our numerical calculation (continuous lines). We can see an excellent agreement between analytical and numerical results. However, the numerical result contains a high-frequency oscillation not seen in the Rabi solution. The reason is that right from the start, the Rabi solution considers only frequencies near resonances, as it assumes a field $V_{int}(t)=V_{int}(0)e^{i2\Omega t}$, while our numerical solution contains resonant and non-resonant frequencies.

By changing $\alpha$ we move away from the solvable three-level system. Figure 
\ref{Fig:ProbabilitiesThreeRabiA}(b) shows that the main effect is a change of $\Omega_w$ as $w$ depends on $\alpha$. The other effect is a reduction of the highest-band occupancy as the FB band is not empty at half-period.

In twisted bilayer graphene and in the $\alpha-T_3$ graphene model, the Fermi level falls at the flat band. Therefore it is relevant to study the system when a pure FB state is taken as initial condition. Interestingly, in this case the three-level Rabi system mimicks a two-level system due to the symmetry. This behavior is seen in Fig. \ref{Fig:ProbabilitiesThreeRabiB} panels (a) and (b). Panel (a) corresponds to the known solvable case $\alpha=1$ in which the numerical
and analytic solution is almost the same. As we change $\alpha$, the period is modified, small ripples are seen due to non-resonant contributions and the highest bands are never fully occupied. 
This suggests that upon charge doping, the system can be changed from a two to a three-level Rabi problem.

\subsection{Resonance frequency $\omega\approx\Omega/2$: Two-level Rabi problem}

\begin{figure}[t]
\begin{center}
\includegraphics[scale=0.65]{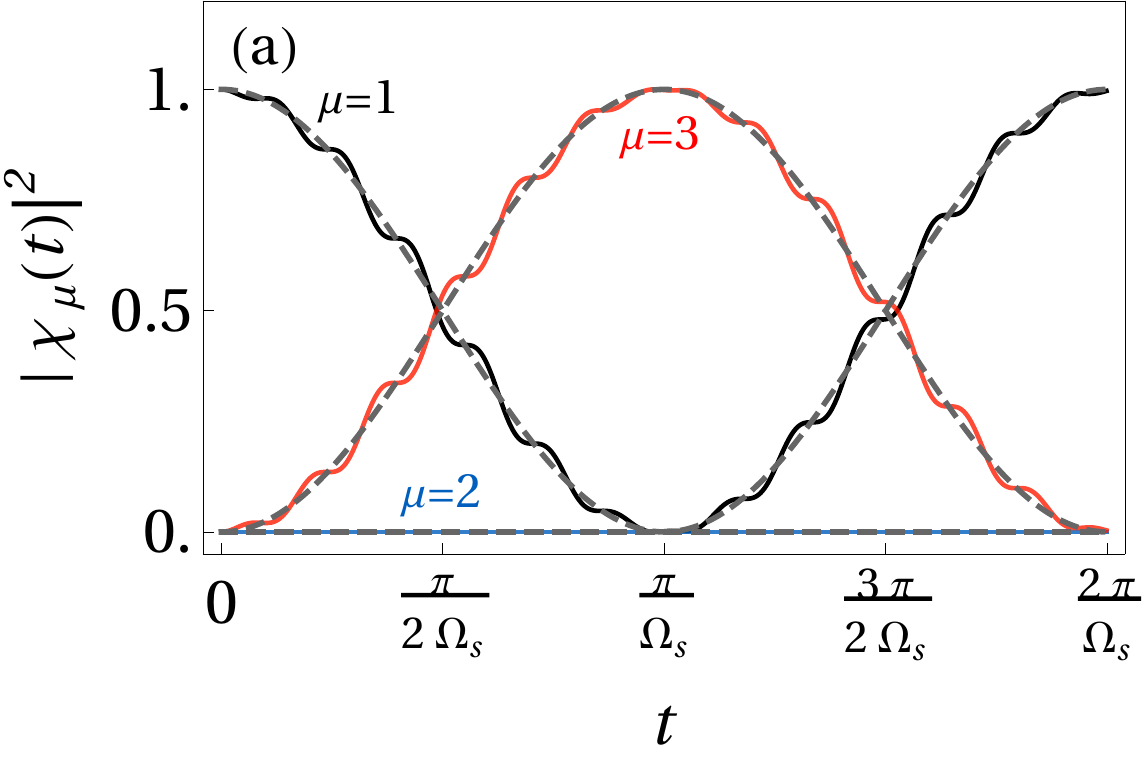}
\includegraphics[scale=0.65]{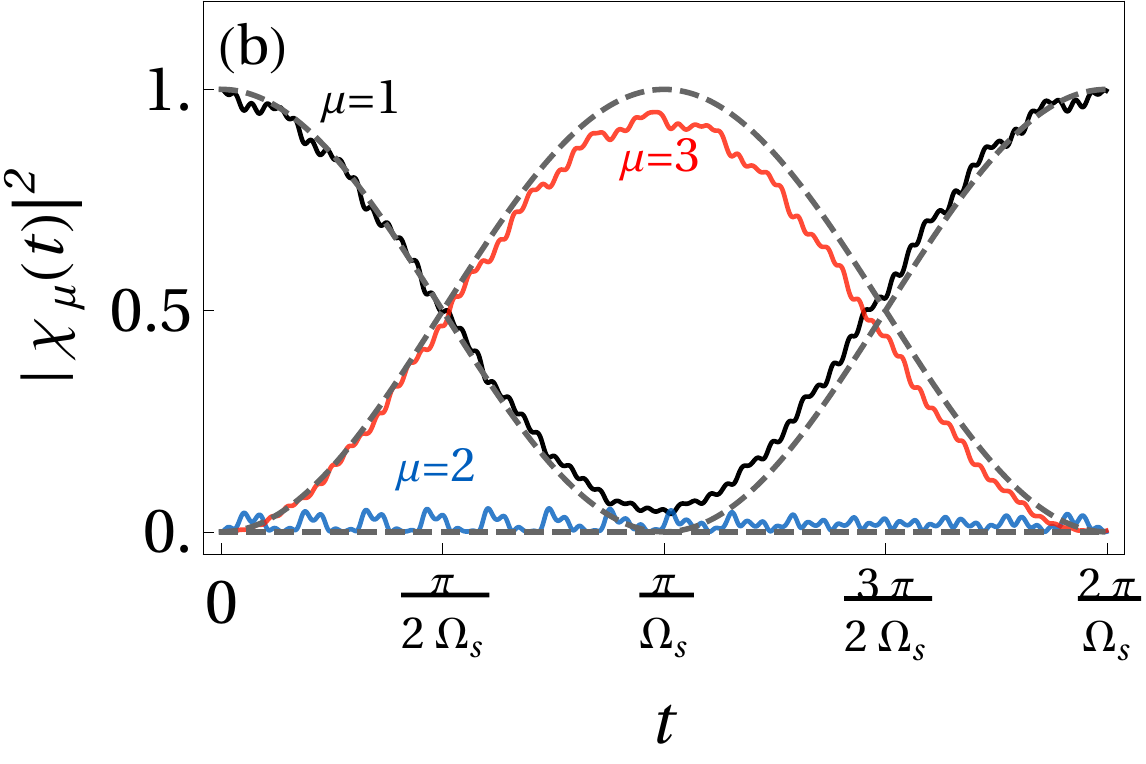}
\end{center}
\caption{\label{Fig:ProbabilitiesTwoRabi} Occupation probabilities $|\chi_{1}(t)|^2$ (black line),  $|\chi_{2}(t)|^2$ (red line) and $|\chi_{3}(t)|^2$ (blue line) as a function of time for the resonant frequency $\omega\approx\Omega/2$. The dotted gray lines are the results obtained for the two-level Rabi problem considering only resonant terms. In, (a) shows the numerical results for $\alpha=0.0$ and $\Omega_s=3.45$ Gz. In, (b) shows the behavior for $\alpha=0.8$ with $\Omega_s=0.75$ Gz.  The initial state in both panels are given by $\chi_{1}(0)=1$. Note that although $\Omega_{s}$ is different in both panels, the periodicity is determined by $\Omega_{s}$. Notice how the case $\alpha=0$ corresponds to the two-level Rabi problem.
}
\end{figure}

As we found in subsection \ref{Subsection:perturbation}, there is a second resonant frequency at $\omega\approx\Omega/2$. In Fig. \ref{Fig:ProbabilitiesTwoRabi}(a) we plot the occupation probabilities for the VB (black line), FB (blue) and CB (red line) for $\alpha=0$, when the initial state is $\chi_{1}(0)=1$. In this case, $\mathcal{C}_\alpha=1$ and $\mathcal{S}_\alpha=0$, and hence the $\alpha-\mathcal{T}_{3}$ lattice is a honeycomb lattice resembling mono-layer graphene \cite{malcolm2016frequency}. 

Hence, $w=0$ and the matrix $\mathbbm{B}(t)$ in Eq. \eqref{ec:MatrixBLinearlyPolarized} is the same of a know simplified solvable form of the two-level Rabi system\cite{RabiTwoLevel1937, dittrich1998quantum}. The band-ocuppances are given by,
\begin{eqnarray}\label{eq:ThreeRabi}
|\chi_{1}(t)|^2&=&\left[\cos\left(\frac{\Omega_s}{2} t\right)\right]^2,\label{eq:TwoRabiOne}\\
|\chi_{2}(t)|^2&=&0,\label{eq:TwoRabiTwo}\\
|\chi_{3}(t)|^2&=&\left[\sin\left(\frac{\Omega_s}{2} t\right)\right]^2\label{eq:TwoRabiThree}.
\end{eqnarray}
The band-occupation presents oscillations, where the period for the BV and BC is $T=2\pi/\Omega_s$ and $\Omega_s=|s|$. In  this particular case $\Omega_s=3.45\,$GHz. The above expressions are in agreement with the time perturbation theory presented before, where  $\mathcal{P}^{A}_{2\leftarrow 1}=\mathcal{P}^{A}_{3\leftarrow 2}=0$ and $\mathcal{P}^{A}_{3\leftarrow 1}\neq0$. Such result indicates that the FB does not contribute to the occupancy probabilities.  
In Fig. \ref{Fig:ProbabilitiesTwoRabi}(a), we show that the numerical solution and the band occupancies (gray dotted lines), given by Eqs. \eqref{eq:TwoRabiOne}-\eqref{eq:TwoRabiTwo} showing an excellent agreement. As in the problem of three-level, the numerical result contains a high-frequency oscillation not seen in the two-level Rabi solution.

Finally, by increasing the value of $\alpha$, the two-level system is difficult to solve and the numerical solution present two effects. As show in Fig. \ref{Fig:ProbabilitiesTwoRabi}(b), the first effect is a reduction of the highest-band occupancy in the BV and CB, and the second is a change in the frequency of $\Omega_s$ due to the shift of $\alpha$ in Eq. \eqref{ec:s_coeficient}. 

\section{Conclusions}\label{Sec:Conclusions}

We investigated the occupancy probabilities for the $\alpha-\mathcal{T}_3$ system under linearly polarized light by studying the corresponding Dirac equation in the interaction picture.  The transitions 
between states have contributions that depend upon the relative angle between the electron momentum and the electromagnetic field wave-vector.  When both are parallel (or antiparallel), the transitions are found by using Floquet theory; while for other directions we used  time-perturbation theory and numerical solutions. This allowed us to found two resonant frequencies. The first resonance $\omega\approx\Omega$, involves the flat band as an intermediate step (three-level system). In contrast, the second, $\omega\approx\Omega/2$ is similar to the valence and conduction band transitions observed in graphene (two-level system). The value of the parameter $\alpha$ plays an important role in the behavior of transitions probabilities. While for $\alpha$ close to one and $\omega\approx\Omega$ the system recovers the behavior of the three-level Rabi system, for $\alpha$ near zero and $\omega\approx\Omega/2$ it behaves like a two-level a Rabi system.
Moreover,  we also showed that upon charge doping, the system can be changed from a two to a three-level Rabi problem, unveiling many interesting possibilities for quantum control and electronic devices\cite{Terrones}.

\section{Acknowledgements}\label{sec:Acknowledgements}

We thank UNAM DGAPA-PROJECT IN102620. V.G.I.S and J.C.S.S. acknowledge the total support from 
DGAPA-UNAM fellowship. M.A.M. thanks to AMC for the support during the XXIV Summer of Scientific Research (2019). R.C.-B.  acknowledges  the  hospitality  of  the  Instituto de F\'{i}sica at UNAM where this project was finished.
\bibliography{Biblio}

\end{document}